\PassOptionsToPackage{table,xcdraw}{xcolor}

\documentclass[sigconf]{acmart}

\AtBeginDocument{%
  }

\copyrightyear{2026}
\acmYear{2026}
\setcopyright{cc}
\setcctype{by}
\acmConference[ICSE '26]{2026 IEEE/ACM 48th International Conference on Software Engineering}{April 12--18, 2026}{Rio de Janeiro, Brazil}
\acmBooktitle{2026 IEEE/ACM 48th International Conference on Software Engineering (ICSE '26), April 12--18, 2026, Rio de Janeiro, Brazil}
\acmPrice{}
\acmDOI{10.1145/3744916.3773266}
\acmISBN{979-8-4007-2025-3/2026/04}

\usepackage[normalem]{ulem}
\usepackage{macros}
\usepackage{multirow}
\PassOptionsToPackage{hyphens}{url}
\usepackage{hyperref}
\usepackage{xurl}

\usepackage{algorithm}
\usepackage[noend]{algpseudocode}

\usepackage{mdframed}

\usepackage{enumitem}

\usepackage{array}

\usepackage{xspace}

\usepackage{cleveref}
\usepackage{hhline}

\usepackage{graphicx}
\usepackage{subcaption}

\usepackage{pifont}
\newcommand{\cmark}{\ding{51}}%
\newcommand{\xmark}{\ding{55}}%

\newcommand{\halfmark}{%
  \ooalign{\raisebox{0.1ex}{\ding{51}}\cr\raisebox{-0.1ex}{\ding{55}}}%
}

\usepackage[table]{xcolor}
\usepackage{soul}

\usepackage{booktabs} 
\usepackage{makecell}
\usepackage{mathtools}

\usepackage{pgf}
\usepackage{tikz}
\usetikzlibrary{tikzmark}
\usetikzlibrary{calc,arrows,automata,fit}
\usetikzlibrary{positioning,shapes,shadows}
\usetikzlibrary{decorations.pathmorphing,snakes}

\usepackage{etoolbox} 

\newcommand{\revise}[1]{\textcolor{black}{#1}}

\newif\ifshowrevisions
\showrevisionstrue  

\definecolor{revisioncolor}{RGB}{0,0,255} 

\newcommand{\revisionhl}[1]{%
  \ifshowrevisions%
    \textcolor{black}{#1}%
  \else%
    #1%
  \fi%
}

\begin{document}


\newcommand*{\name}{{\textsc{CrossGuard}}\xspace}

\newcommand*{\tool}{{\textsc{Trace2Inv}}\xspace}



\title{Enforcing Control Flow Integrity on DeFi Smart Contracts}

\author{Zhiyang Chen}
\orcid{0000-0002-2315-397X}
\affiliation{%
  \institution{University of Toronto}
  \city{Toronto}
  \country{Canada}
}
\email{zhiychen@cs.toronto.edu}

\author{Sidi Mohamed Beillahi}
\orcid{0000-0001-6526-9295}
\affiliation{%
  \institution{University of Toronto}
  \city{Toronto}
  \country{Canada}
}
\email{sm.beillahi@utoronto.ca}

\author{Pasha Barahimi}
\authornote{Pasha is an incoming Ph.D. student at the University of Southern California.
Work done while the author was a remote research intern at the University of Toronto.}
\orcid{0009-0005-6803-0438}
\affiliation{%
  \institution{University of Tehran}
  \city{Tehran}
  \country{Iran}
}
\email{pashabarahimi@gmail.com}

\author{Cyrus Minwalla}
\orcid{0000-0002-9569-664X}
\affiliation{%
  \institution{Bank of Canada}
  \city{Ottawa}
  \country{Canada}
}
\email{CMinwalla@bank-banque-canada.ca}

\author{Han Du}
\orcid{0009-0005-0256-180X}
\affiliation{%
  \institution{Bank of Canada}
  \city{Ottawa}
  \country{Canada}
}
\email{HDu@bank-banque-canada.ca}

\author{Andreas Veneris}
\orcid{0000-0002-6309-8821}
\affiliation{%
  \institution{University of Toronto}
  \city{Toronto}
  \country{Canada}
}
\email{veneris@eecg.toronto.edu}

\author{Fan Long}
\orcid{0000-0001-7973-1188}
\affiliation{%
  \institution{University of Toronto}
  \city{Toronto}
  \country{Canada}
}
\email{fanl@cs.toronto.edu}


\begin{abstract}

Smart contracts power decentralized financial (DeFi) services but are vulnerable to
security exploits that can lead to significant financial losses. 
Existing security measures often fail to adequately protect these contracts due to the
composability of DeFi protocols and the increasing sophistication of attacks. 
Through a large-scale empirical study of historical transactions from the 
\revise{$37$} hacked DeFi protocols, we discovered 
that while benign transactions typically exhibit a limited number of unique 
control flows, in stark contrast, attack transactions consistently 
introduce novel, previously unobserved control flows.
Building on these insights, we developed {\name}, a novel framework that enforces control 
flow integrity onchain to secure smart contracts. Crucially, {\name} does not 
require prior knowledge of specific hacks. Instead, configured only once 
at deployment, it enforces control 
flow whitelisting policies 
and applies simplification heuristics at runtime. 
This approach monitors and prevents potential attacks by reverting all transactions 
that do not adhere to the established control flow whitelisting rules.
Our evaluation demonstrates that {\name} effectively blocks \revise{$35$} of the 
\revise{$37$} analyzed 
attacks when configured only once at contract deployment, maintaining a low false positive 
rate of \revise{$0.26\%$} and minimal additional gas costs. These results underscore the efficacy 
of applying control flow integrity to smart contracts, significantly enhancing security 
beyond traditional methods and addressing the evolving threat landscape in the DeFi ecosystem.

\end{abstract}


%





\maketitle

\vspace{-2ex}
\section{Introduction} 
\label{sec:introduction}

Blockchain technology has revolutionized the creation of global, secure, and
programmable ledgers, fundamentally altering how digital transactions are
conducted. Central to this innovation are smart contracts, which operate on
blockchains, allowing developers to define and enforce complex
transactional rules directly on the ledger. This capability has positioned
smart contracts as the backbone of various decentralized
financial (DeFi) services. As of March 13th, 2025, the total value locked in
3,973 DeFi protocols has surged to approximately \$87.82
billion~\cite{defillama}, highlighting the substantial economic impact and
growth of this technology.

However, by the same date, vulnerabilities in DeFi smart contracts have resulted in 
financial losses exceeding \$11.21 billion USD~\cite{defillamahacks}. 
In response, researchers have developed program analysis and verification techniques 
to secure smart contracts~\cite{luu2016making, torres2018osiris, 
liu2018reguard, feist2019slither, zhang2019mpro, bose2022sailfish, 
ma2021pluto, zheng2022park, wang2024efficiently, albert2020taming, 
zhang2023your}, and developers often commission security audits prior to deployment.


Despite these advancements in security measures, the evolving landscape of smart 
contracts has continually outpaced traditional defenses.
Modern smart
contracts are now designed with the flexibility to support the layering of
additional contracts, a feature particularly vital in DeFi ecosystem. In DeFi, smart contracts facilitate a diverse array of
financial products and services, such as lending and
yield farming. These interlinked contracts are often referred to as ``DeFi
legos,'' emphasizing their modularity. The ability to combine various DeFi smart
contracts, a concept known as ``DeFi composability,'' is widely regarded as one
of the key advantages of DeFi~\cite{popescu2020decentralized, amler2021defi,
schar2021decentralized}. 
However, this complexity and interdependence introduce
significant challenges to securing smart contracts with conventional methods.
The security of a DeFi protocol depends not only on the correct design and 
implementation of its own contracts but also on the integrity of external 
contracts it interacts with. Additionally, experienced users or attackers can 
deploy their own smart contracts to invoke functions across 
multiple DeFi protocols in arbitrary sequences. 
Considering all possible interactions a DeFi 
protocol may encounter prior to deployment is infeasible for traditional 
security approaches.

A critical observation in hack transaction analysis is that they often
\emph{exploit unintended control flows across multiple functions}, deviating
from the original design intentions of the developers. For instance,
re-entrancy attacks leverage an unforeseen recursive control flow through
default handlers in custom contracts, enabling repeated execution of a critical
function within one transaction. Similarly, flash loan attacks manipulate
multiple functions in a precisely timed sequence, utilizing large asset
transfers to coerce the victim contract into executing unfavorable trades.
We conducted an empirical study analyzing transaction histories of 
\revise{$37$} compromised Ethereum protocols. 
Our findings reveal that control flows are relatively constrained, 
with attack transactions introducing novel flows never observed 
previously in all but one hacked protocol.



\noindent \textbf{{\name}:} 
Building on the above observations, we developed {\name}, a novel framework to
enforce \emph{control flow integrity} to secure smart contracts. Given a DeFi
protocol, {\name} instruments its existing smart contracts with additional code
to track control flow data. {\name} also deploys a new guard contract that
collects control flow data from these instrumented contracts, and enforces four
whitelisting policies at runtime to detect and neutralize any attacks that
attempt to exploit unexpected control flows. Unlike many previous invariant
enforcement tools~\cite{chen2024demystifying, liu2022invcon}, {\name} does not
rely on inferring its security rules from prior benign transaction traces and
therefore can apply to smart contracts immediately at their initial
deployments, leaving no gap of unprotected periods.

A key challenge arises from {\name}'s whitelisting-only design. 
This approach, while enhancing security by adhering
strictly to known safe paths, could inherently lead to an increase in false
positives if not meticulously managed. Although the number of unique control
flows is inherently limited, a naive whitelisting approach could still produce
numerous false positives or demand substantial human intervention. To address
this issue,  {\name} employs heuristics to simplify control flows: 
excluding read-only calls that don't alter state, 
tracking read-after-write dependencies, and treating independent calls as 
separate flows. These heuristics effectively simplify control flows,
and lower false positive rates.


\noindent \textbf{Experimental Results:}
We evaluated {\name} on the deployed smart contracts and their transactions 
of the \revise{$37$} hacked DeFi protocols
included in our empirical study. Our results indicate 
that, when 
configured only once at deployment, {\name} can effectively prevent 
\revise{$35$} out of \revise{$37$}
attacks analyzed in our study, maintaining a low average false positive rate of
just \revise{0.26\%}. Unlike traditional methods, {\name} does not depend on historical
transactions. Despite this, {\name} still surpasses the state-of-the-art which 
instruments the smart contracts with invariants learned from historical transactions.
Moreover, after implementing \revise{four}
optimization techniques, {\name} achieves a minimal gas consumption overhead of
\revise{3.53\%} on average. These results demonstrate the usefulness of our empirical
findings and {\name}. 

\noindent\textbf{Contributions:} This paper makes the following contributions. 
\begin{itemize}[leftmargin=*]
    \item \textbf{Empirical Study:} To the best of our knowledge, we conducted
        the first comprehensive empirical study of control flows in historical
        transactions of compromised DeFi protocols. Our analysis uncovers
        critical insights into the control flow patterns prevalent in DeFi protocols and
        explores various use cases of DeFi composability.

    \item \textbf{{\name} Technique:} This paper proposes the first control flow integrity
        technique for smart contracts with whitelisting policies and
        simplification heuristics. This paper also details methods for implementing 
        these policies and heuristics through static and dynamic analysis, and describes
        how they are instrumented in contracts and enforced on the fly. 

    \item \textbf{Evaluation and Tools:} This paper evaluates the effectiveness
        of {\name} in preventing attacks. To support ongoing research and
        facilitate community engagement, we provide open access to the
        study results, experimental results, and our tool,
        available at our website~\cite{website}.
\end{itemize}
\vspace{-2mm}

\vspace{-2ex}
\section{Background and Empirical Study}
\label{sec:study}

The \textbf{Ethereum Virtual Machine (EVM)} 
executes smart contracts and maintains network state. 
\textbf{Gas} measures computational effort for transaction execution, 
with storage operations being particularly expensive. 
\revise{The recent \textbf{TLOAD} and \textbf{TSTORE} 
opcodes~\cite{eip1153,kraken2024ethereum} provide temporary 
transaction-scoped storage at reduced gas costs, 
with variables automatically initialized to zero at transaction start.}
\textbf{DeFi protocols} are decentralized financial systems using 
interconnected smart contracts for lending, borrowing, and trading. 
\textbf{Control Flow} in software engineering is the order in 
which program instructions execute. In smart contracts, 
runtime control-flow analysis can be performed at different granularities, 
from low-level opcodes to high-level function calls, 
raising the question of which level offers the best 
trade-off between security coverage and computational cost. 

\begin{figure*}[!h]
\centering
\includegraphics[width=0.8\textwidth]{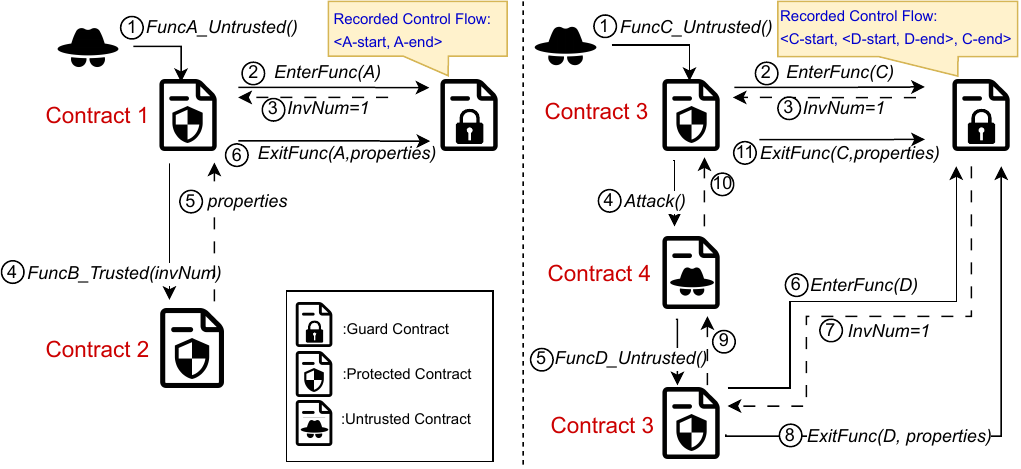}
    \vspace{-2mm}
    \caption{Running example showing \textsc{CrossGuard}'s response to simple invocation 
    (left) and re-entrancy attack (right)}
    \Description{Two-panel running example. Left: a benign/simple invocation sequence. Right: a re-entrancy attack sequence. The figure illustrates how \textsc{CrossGuard} observes function-level control flow and reacts differently in benign versus attack cases.}
    \label{fig:example}
\end{figure*}

To ground our framework, we hypothesize that \emph{function-level} 
control flows are sufficient to distinguish malicious from 
benign transactions, while remaining practical overheads
for runtime monitoring. 
To evaluate this hypothesis, we analyze each 
victim protocol's 
transaction history prior to the incident. 
Because protocols typically 
operate for some time 
before being exploited and attract many 
interacting actors, benign behavior can 
evolve over time, diversifying and introducing 
previously unseen control flows. 
This motivates the following research question:

\noindent \textbf{RQ1: 
How do \revise{(function level)} control flows in hack transactions differ 
from those in other 
(benign) transactions 
prior to a hack?}

To answer RQ1, we conducted a study on a systematically collected 
benchmark comprising \revise{$37$} victim protocols involved in security 
hacking incidents. \revise{The scope of this work is described in 
Section~\ref{sec:preli}, and the details of our benchmark collection 
methodology are provided in Section~\ref{subsec:study_methodology}.} 
For each hacking incident, we examine the uniqueness 
of the control flows in the hack transactions, determining whether 
they differ from all 
previously observed transactions. The detailed study results are available 
in the ``RQ1'' column in Table~\ref{tab:studyTable} in 
Section~\ref{sec:evaluation}. 
In this study, we only collect the top-level function calls 
to contracts of the victim protocol, 
ignoring deeper control flows within external calls.



\noindent \textbf{Results:}
Out of the \revise{$37$} studied hack incidents, 
\revise{$32$} demonstrated control flows that
were distinct from any previously observed transaction patterns (marked as
\cmark in Table~\ref{tab:studyTable}), highlighting the novel mechanisms by
which these exploits were conducted. However, in $5$ cases (bZx, VisorFi, 
Opyn, DODO and
Bedrock\_DeFi), the control flows had been observed previously in benign
transactions. A detailed investigation into these exceptions revealed
insightful nuances. In particular, the Bedrock\_DeFi hack
(marked as \xmark) was traced back to an issue which mistakenly set the
conversion ratio of ETH/uniBTC to 1:1. This exploit was executed
through one function call to steal funds, which, while not novel in terms
of control flow pattern, leveraged a specific code vulnerability to 
breach the
protocol~\cite{bedrock2025}. Catching this hack involves
combining control flow analysis with data flow analysis.
The remaining $4$ hacks posed a different complexity: 
each involved various
types of re-entrancy (marked as $\halfmark$). 
Although these attacks appeared to have familiar top-level 
function calls, they triggered unique and sophisticated control 
flows at deeper interaction levels. 
This shows that, even the
simple control flow-based analysis 
can detect \revise{$32$} hacks,
it misses $4$ that require a more refined control 
flow-based approach to identify. 
Thus, we propose to enhance the control flow analysis 
framework 
to more accurately detect and classify attacks with 
intricate control flows, 
such as these $4$ re-entrancy hacks. 
The details of this enhanced 
approach are presented in Section~\ref{sec:approach}. 
Unless otherwise stated, 
we use the term ``control flow'' throughout 
the rest of this paper to refer to function-level control flow.

\noindent \textbf{Finding:} 
the vast majority of hack transactions introduce
unique function-level control flows that differ from all 
previously observed benign transactions.

\section{Running Example}
\label{sec:example}




To illustrate how \name monitors and enforces control flow 
integrity, we present two contrasting scenarios: a benign transaction 
with simple invocation and a malicious re-entrancy attack. 
Figure~\ref{fig:example} demonstrates {\name}'s response to these 
two different 
transactions.
The left side of Figure~\ref{fig:example} illustrates a benign transaction 
involving two protected contracts. Contract 1 contains \texttt{FuncA},
whose code externally invokes \texttt{FuncB} in Contract 2. 
When a user initiates this interaction, 
because they are not part of the protected protocol, 
they can only access 
\texttt{FuncA\_Untrusted()}, which triggers the monitoring mechanism.
Contract 1 first calls \texttt{EnterFunc(A)} to notify the guard contract of 
function entry, receiving an invocation number \texttt{invNum=1} that 
uniquely identifies this execution count. 
Since \texttt{FuncB} is an 
expected external call within \texttt{FuncA}, and both contracts 
belong to the protected protocol, Contract 1 invokes 
\texttt{FuncB\_Trusted(invNum)} directly, 
bypassing additional guard contract 
interactions while still passing the 
invocation number for runtime property tracking.
Contract 2 monitors its execution properties and returns them to 
Contract 1, which then calls \texttt{ExitFunc(A, properties)} to report completion 
along with collected runtime properties. The guard contract ultimately 
records a simple control flow pattern \texttt{<A-start, A-end>}, which 
represents normal protocol behavior and is classified as benign.

\revise{
The right side of Figure~\ref{fig:example} demonstrates how {\name} 
stops a cross function re-entrancy attack. 
An attacker deploys Contract 4 and exploits Contract 3, 
which contains \texttt{FuncC} and \texttt{FuncD} that 
together constitute a re-entrancy vulnerability. 
The attack begins when 
the attacker invokes \texttt{FuncC\_Untrusted()}, 
which subsequently triggers
\texttt{EnterFunc(C)} 
and receiving \texttt{invNum=1}. 
During execution, Contract 3 invokes 
\texttt{Attack()}
in the attacker's Contract 4, 
which then re-enters Contract 3 and calls 
\texttt{FuncD\_Untrusted()}. 
This triggers another \texttt{EnterFunc(D)} call, which returns the same 
\texttt{invNum=1}, indicating that this function call occurs within the context of 
the ongoing invocation. When both functions complete their execution, 
they call their respective \texttt{ExitFunc} methods, resulting in the 
guard contract recording 
a nested control flow pattern \texttt{<C-start, <D-start, D-end>, C-end>}. 
This nested structure clearly indicates 
a same-contract cross function
re-entrancy behavior, and 
unless explicitly whitelisted by protocol administrators, 
{\name} blocks such transactions by default.}

\revise{
The example highlights several key aspects of {\name}'s design. 
The dual function variants(defined in Section~\ref{sec:preli}) 
approach optimizes performance 
by allowing trusted intra-protocol calls to bypass guard contract 
interactions and save gas. 
The invocation numbering system enables correlation of related 
function calls within complex control flow sequences. 
Critically, both function variants collect runtime properties 
that are passed to the guard contract, providing additional context 
beyond control flow patterns to reduce false positives. 
The guard contract uses both control flow structure and 
runtime properties to distinguish between 
legitimate sequential execution patterns and malicious anomalies.
}

\section{Definitions, Threat Model and Scope}
\label{sec:preli}




We now formalize the control flow of a transaction 
which \name relies on. 
We use \textbf{protected protocol}, denoted as $P$, to refer 
to the set of smart contracts protected by our technique. 
We use $C_i$ to represent the $i^{th}$ \textbf{protected contract}, 
where $P = \{ C_1, C_2, \dots, C_j \}$. Typically, 
$P$ consists of all the core contracts of the protocol, 
as specified by the developers of the protocol. 
We use \textbf{external contracts}, denoted as $E$, to refer 
to smart contracts that a protected protocol is built on top of 
but are external to the protected protocol, such as stable coins or oracles. 
In addition, any contract that is neither part of the protected 
protocol nor considered external is referred to as an 
\textbf{untrusted contract} (denoted as $U$). All functions 
within protected contracts are referred to as
\textbf{protected functions}.


\begin{definition}[Tracked Runtime Properties]
As shown in Section~\ref{sec:example}, {\name} tracks runtime properties
for each execution of a protected function, with the detailed algorithms
given in Section~\ref{subsec:algorithm}. Specifically, it tracks the
following two properties:
\begin{itemize}[leftmargin=*, itemsep=0pt, topsep=0pt, partopsep=0pt]
\item \textbf{Runtime Read-Only} ($\mathit{isRR}$): 
  True if the function does not change any blockchain state on the fly.
\item \textbf{Read-After-Write Dependency} ($\mathit{isRAW}$):
  True if, within the same invocation count (as tracked by $\mathit{invNum}$), 
  the function reads from any storage location that was written by a 
  previous invocation in the same transaction.
\end{itemize}
Both properties are maintained as boolean flags and are 
\emph{mergeable} under 
intra-protocol composition: for a protected function $f$ calling another
protected function $g$ within the same invocation, 
$f$ aggregates 
$g$'s runtime properties upon return to reflect the combined behavior.
The merged properties 
at return are updated as $\mathit{isRR}_f \coloneqq \mathit{isRR}_f 
\land \mathit{isRR}_g,\quad \mathit{isRAW}_f \coloneqq 
\mathit{isRAW}_f \lor \mathit{isRAW}_g.$
\end{definition}

{\name} instruments each 
protected contract function with two variants to optimize monitoring 
overhead while maintaining security guarantees.
For each function $f$ in a protected contract $C_i \in P$, 
\textsc{CrossGuard} creates two variants. Both variants collect
and propagates runtime properties.
\begin{itemize}[leftmargin=*, itemsep=0pt, topsep=0pt, partopsep=0pt]
\item \textbf{Untrusted variant} $f\_\text{Untrusted}$: Accessible to any 
caller but mandatory triggers \texttt{EnterFunc} and \texttt{ExitFunc} 
of guard contract.
\item \textbf{Trusted variant} $f\_\text{Trusted}$: Restricted to calls 
from other protected contracts via access control, 
bypasses guard contract interactions. 
\end{itemize}



\begin{definition}[Call Tree and Invocation]
The \textbf{call tree} $CT(tx)$ of transaction $tx$ represents the tree 
structure of function calls, where nodes correspond to function calls 
and directed edges represent call dependencies. 
An \textbf{invocation} $\iota(tx, P)$ is a sequence of 
function calls starting from an entry point in protected protocol $P$. 
Invocations are classified as \textbf{re-entrant} (containing 
recursive calls to protected contracts
through untrusted intermediaries), or
\textbf{simple} (no re-entrancy to protected contracts).
\end{definition}


\begin{definition}[Control Flow]
The \textbf{control flow} of an invocation $CF(\iota)$ 
captures the 
start and end of function calls within protected contracts, 
abstracting away 
calls to external contracts unless they facilitate re-entrancy to 
protected contracts. The \textbf{control flow of a transaction} 
$CF(tx, P)$ is defined as the sequence 
 of 
all invocation control flows $\langle CF(\iota_1),\ldots, CF(\iota_n) \rangle$. A control flow is 
\textbf{trivial} only if it consists of a simple 
invocation, otherwise it is \textbf{non-trivial}.
\end{definition}

\revise{
Consider the re-entrancy scenario in Section~\ref{sec:example}. 
The two invocations would be $\iota_1$ 
(the initial \texttt{FuncC} call) and $\iota_2$ (the re-entrant \texttt{FuncD} call). 
The resulting control flow $CF(tx, P) = \langle \langle C_3.\texttt{C-s}, 
\langle C_3.\texttt{D-s}, 
\\ C_3.\texttt{D-e} \rangle, C_3.\texttt{C-e} \rangle \rangle$ clearly reveals 
re-entrancy, where notation `-s' and `-e' denote function start and end respectively.
}

\smallskip
\noindent \revise{\textbf{Threat Model: }}
\revise{
Our threat model assumes sophisticated attackers who can deploy 
arbitrary untrusted contracts to interact with protected protocols. 
Given the transparent nature of blockchain systems, 
attackers have access to {\name}'s on-chain implementation, 
enabling them to probe its detection mechanisms 
and whitelisting 
policies/heuristics.
Additionally, attackers may attempt to evade detection by 
decomposing complex attacks into multiple simpler transactions 
or adapting their strategies to minimize detection risks 
while maximizing profit. We also assume attackers do not have 
administrative 
access to the protected protocol.}

\noindent \revise{\textbf{Scope: }}
\revise{
{\name} blocks attacks that exhibit non-trivial 
control flows(those utilizing multiple function invocations), 
or re-entrancy patterns. 
{\name} cannot prevent vulnerabilities exploitable 
through single function calls with trivial control flows, 
such as access control bypasses or bridge exploits.}

\section{Approach} 
\label{sec:approach}

In this section, we present the design of {\name}.
We begin by introducing four whitelisting policies, followed by 
two heuristics aimed at reducing false positives.
Finally, we describe how these policies and heuristics are 
implemented within a smart contract, and how 
they enable real-time control flow tracking and simplification.



\vspace{-1ex}
\subsection{Control Flow Whitelisting Policies}
\label{subsec:policy}
\vspace{1sp}

We propose four control flow whitelisting policies.  

\noindent \textbf{Policy 1: Simple Independent Invocations.}
An invocation in a transaction is considered benign if it is \emph{simple} and independent 
of any prior invocations executed within the transaction. The rationale for this policy is that a simple, 
independent invocation mirrors the behavior of a function being invoked by an EOA in a single, 
standalone call. This represents the fundamental usage of a 
function~\footnote{If a simple invocation is flagged as malicious, 
it points to a single function access control issue, 
which falls outside the scope of this paper.}. As such, this type of invocation is considered benign.

\noindent \textbf{Policy 2: Read-Only Invocations.}
An invocation is considered \textbf{benign} if it is \emph{read-only} 
and does not modify any blockchain state. Specifically, an invocation is 
read-only if it performs no write operations to the storage (i.e., no 
\emph{SSTORE} operations), does not call other state-changing functions, and
does not transfer Ether~\footnote{Other operations, such as 
\emph{SELFDESTRUCT} or \emph{CREATE}, 
could also alter the blockchain state but are rare in high-profile DeFi protocols. 
If such operations are present in a function, that function should not be 
classified as read-only.}. This is because invocations not altering
blockchain state have no effect on the final state. 
Therefore, those invocations can be 
omitted from the transaction without influencing the overall outcome.

\noindent \textbf{Policy 3: Runtime Read-Only (RR) Function Calls.}
A function call is considered \textbf{runtime read-only} if it does not 
have a storage write (SSTORE), and it performs no Ether transfers. 
Some functions may not be marked as read-only in their source code 
but behave as such on the fly. 
An invocation is runtime read-only if all function calls 
within it are runtime read-only. 
By tracking runtime behavior and identifying such 
function calls and invocations, we can prune these invocations 
from the control flow, thereby simplifying it.

\noindent \textbf{Policy 4: Restore-on-Exit (RE) Storage Writes.}
This heuristic permits to safely ignore storage writes that temporarily alter values but 
restore the original state at the end of execution, i.e., the value returned in the 
first read operation (SLOAD) equals the one written by its last write operation, 
with no write preceding the first read. 
This increases the likelihood of classifying function calls as 
runtime read-only. A typical example is re-entrancy guard, where a 
function restores the original state of the guard before exiting to prevent re-entrancy attacks. 
In our system, while these re-entrancy guards remain active and function as intended, 
any storage writes to them are disregarded when determining whether a function 
call is runtime read-only.

\vspace{-1ex}
\subsection{Control Flow Simplification Heuristics}
\label{subsec:heuristics}
\vspace{1sp}

To reduce the complexity of control flows and minimize false positives, 
we propose two heuristics 
that allow {\name} to safely ignore certain function calls.

\noindent
\textbf{Heuristic 1: ERC20 Function Calls.}
\emph{ERC20}~\cite{ERC20} is a widely used smart contract standard for implementing tokens 
in DeFi protocols. ERC20 contracts perform token management, and several functions within 
these contracts modify user properties without impacting the overall protocol
state. For example, the functions \texttt{transfer} and \texttt{transferFrom}  
change only the balances of the sender and receiver, while \texttt{approve}, \texttt{increaseAllowance}, and \texttt{decreaseAllowance} simply modify the allowance 
granted by the sender to the spender. We consider these five ERC20 
functions to be benign and safe, as they have been extensively tested and are 
widely used across numerous DeFi projects. Therefore, calls to those 
functions within invocations are safely ignored.

\noindent 
\textbf{Heuristic 2: Read-After-Write (RAW) Dependency.}
An invocation \( \iota_2 \) is considered \textbf{storage read-after-write (RAW)-dependent} on an 
earlier invocation \( \iota_1 \) if:
\( \iota_2 \) reads from a storage location that \( \iota_1 \) writes to (with an exception of 
storage writes classified in Policy 4). 
In the absence of such dependencies, 
invocations are treated as independent, 
and a control flow consisting only simple 
and independent invocations is whitelisted. Note 
that re-entrant invocations,
no matter if it is dependent, will never be 
whitelisted by this heuristic.

\vspace{-1ex}
\subsection{Soundness and Limitation Analysis}
\vspace{1sp}


As mentioned in Section~\ref{sec:preli}, 
{\name} is a general-purpose defense designed to
prevent \emph{all} smart 
contract attacks, with the only 
exception of those exploitable via a 
single function call to a protected contract 
(e.g., integer overflow or access control problems).

\revise{
Policies 1-4 are sound by design as they only whitelist 
control flows that preserve security properties: 
Policy 1 ensures whitelisted transactions are equivalent 
to direct EOA calls, Policies 2-3 are inherently safe 
since read-only operations cannot alter blockchain state, 
and Policy 4 maintains soundness by ensuring temporary 
state changes are reverted, still not altering blockchain states.}

\revise{
However, our heuristics operate on a best-effort basis 
with specific limitations. 
Heuristic 1 assumes standard ERC20 functions maintain expected behavior, 
which could be violated by non-standard implementations.
Heuristic 2 tracks read-after-write dependencies only within 
protected contracts and may miss external state manipulations 
such as oracle manipulations.
However, as shown in Section~\ref{sec:evaluation}, this 
limitation impacted only 1 out of 37 benchmarks evaluated 
(i.e. Bedrock\_DeFi).
}



\vspace{-1ex}
\subsection{System Overview}
\label{subsec:algorithm}
\vspace{1sp}

In this section, we explain how {\name} is integrated into a DeFi protocol pre-deployment 
by instrumenting the original code. 
The system consists of two main components: 
instrumentation within the protected contracts and 
a guard contract.
Each function in the protected contracts is instrumented and 
assigned 
a unique positive integer \texttt{funcID} as 
its identifier.

\begin{algorithm}
    \caption{EnterFunc and ExitFunc in guard contract}
    \label{alg:enterexit}
    \begin{algorithmic}[1]
    \State \textbf{State Variables (accessed via tload/tsstore):}
    \State \quad $sum, invCount: int$  
    \State \quad $callTrace: int[]$  
        \State \quad $isCFRAW, isCFReEntrancy: bool$ 
    \State \quad $\_allowedPatterns: mapping(int \to bool)$  

    \Function{EnterFunc}{funcID: int}
        \If{$sum = 0$} 
            \State $invCount \gets invCount + 1$ 
        \Else
            \State $isCFReEntrancy \gets true$ 
        \EndIf    
        \State $sum \gets sum + funcID$ 
        \State $callTrace.push(funcID)$ 
        \State \Return $invCount$
    \EndFunction

    \Function{ExitFunc}{funcID: int, isRR, isRAW: bool}
        \State $sum \gets sum - funcID$ 
        \If{isRR}
            \State $callTrace.pop()$ 
        \Else
            \State $callTrace.push(-funcID)$ 
        \EndIf
        \State $CFHash \gets 0$ 
        \If{$sum = 0$}
            \For{each $id$ in $callTrace$}   
                \State $CFHash \gets \text{keccak256}(id, CFHash)$
            \EndFor
            \State $callTrace.clear()$ 
        \EndIf
        \If{isRAW}
            \State $isCFRAW \gets true$ 
        \EndIf
        \If{$\neg$ $\_allowedPatterns[CFHash] \land (isCFReEntrancy \lor isCFRAW)$}
            \State \textbf{revert ``Unsafe pattern detected''}
        \EndIf
    \EndFunction

    \end{algorithmic}
\end{algorithm}


When an instrumented function is invoked, 
it sends the guard contract its \texttt{funcID}
to 
record its function entry 
and receives a positive integer \texttt{invCount}
indicating the invocation count. 
The instrumented function then tracks storage accesses, 
records storage writes using \texttt{invCount},  
and sends tracked runtime properties to the guard 
contract upon exit.
The guard contract collects control flows and  
tracked runtime properties
from the protected 
contracts, simplifies control flows using 
the defined policies and heuristics, 
evaluates whether the control flow is whitelisted, 
and reverts the transaction if it is not.~\footnote{Note 
when protocol administrators execute a transaction, {\name} 
includes a straightforward
mechanism (not detailed but trivial to implement) 
that allows administrators to deactivate
the guard contract at the beginning of a transaction, 
perform actions without interference from the
control flow integrity checks, and reactivate the 
guard contract at the end.
}
Alg.~\ref{alg:enterexit} outlines the implementation within 
the guard contract, while Alg.~\ref{alg:instrumented} outlines 
the execution of the instrumented 
protected functions. Furthermore, 
Table~\ref{tab:instrumentation} details 
the instrumentation 
applied to the original source code.

\noindent
\textbf{Algorithm in the guard contract.}
Alg.~\ref{alg:enterexit}  implements the control flow tracking in the guard contract. 
The function \texttt{EnterFunc} is invoked by protected contracts 
when one of their functions is called by an untrusted contract, 
taking a unique function identifier (\texttt{funcID}) 
as input for each function in the protected contracts.
If the sum of function identifiers is zero, it signifies the start of a new invocation, 
and \texttt{invCount} is incremented (line 8). 
Otherwise, it indicates a re-entrancy condition, and \texttt{isCFReEntrancy} is set to 
\texttt{true} (line 10). The \texttt{sum} and \texttt{callTrace} are updated (lines 11-12) 
to record \texttt{funcID}. 
The \texttt{EnterFunc} returns \texttt{invCount} to the 
protected contracts, enabling them to 
internally track runtime properties.

The \texttt{ExitFunc} (lines 14-28)~\footnote{
    Both \texttt{EnterFunc} and \texttt{ExitFunc} have access control 
    to check if the caller is a protected contract, which is omitted for 
    brevity in 
    Alg.~\ref{alg:enterexit}.
} is called by the protected 
contracts at the exit of the same function that triggered \texttt{EnterFunc}. 
It takes three arguments: \texttt{funcID}, \texttt{isRR} (isRuntimeReadOnly), and \texttt{isRAW} 
(isRead-After-Write dependent on a previous invocation). 
If the invocation is runtime read-only, it removes the \texttt{funcID} added by \texttt{EnterFunc} 
from the \texttt{callTrace} (line 17). 
Otherwise, it pushes the negated \texttt{funcID} 
(which represents the end of the function call)
onto the stack 
(line 19). When the \texttt{sum} equals zero, signaling the end of an invocation, 
the \texttt{CFHash} is computed over the entire 
\texttt{callTrace} (lines 22-23) to summarize the 
control flow of the invocation as a hash. 
Additionally, if any invocation has 
a RAW dependency, Alg.~\ref{alg:enterexit} 
sets the \texttt{isCFRAW} flag to \texttt{true} (lines 25-26). 
Finally, if the computed hash does not match an allowed pattern, and a read-after-write condition 
or a re-entrancy condition is detected, the transaction is reverted to prevent unsafe behavior, 
enforcing the policy to block malicious control flows (lines 27-28). Without any pre-approved 
control flow patterns, \texttt{\_allowedPatterns} 
only include simple invocations by default. But administrators
can add more patterns to this mapping to whitelist more control flows.

\begin{algorithm}
    \caption{State Access Tracking in Protected Functions (Each step within 
    in this algorithm is executed as part of the instrumented code, in 
    accordance with the modifications outlined in Table~\ref{tab:instrumentation}.)}
    \label{alg:instrumented}
    \begin{algorithmic}[1]
    \State \textbf{State Variables (accessed via tload/tsstore):}
    \State \quad $storageWrites: \text{mapping(mapping(bytes $\to$ int) $\to$ bool)}$ 
    \State \quad $tempReads, tempWrites: \text{mapping(bytes $\to$ bytes)}$ 

    \Function{ExecuteInstrumentedCode}{invNum: int}
    \State $readElements, writeElements: \text{arrays of int} \gets []$ 
    \State $isRR: \text{bool} \gets \text{true}$   
    \State $isRAW: \text{bool} \gets \text{false}$   

    \State \textbf{Execute the original source code with instrumentation outlined in 
    Table~\ref{tab:instrumentation}}.

    \For{each $slot$ in $writeElements$} 
        \State $storageWrites[invNum][slot] \gets true$  
        \If{$tempWrites[slot] \neq tempReads[slot]$}  
            \State $isRR \gets false$             
        \EndIf
    \EndFor

    \For{each $slot$ in $readElements$}                 
        \For{$i \gets 1$ to $invNum - 1$}
            \If{$storageWrites[i][slot]$}    
                \State $isRAW \gets true$
                \State \textbf{break} 
            \EndIf
        \EndFor
    \EndFor
    \State clear $tempReads$ and $tempWrites$
    \State \textbf{return} $isRR, isRAW$ 
\EndFunction

\Function{FuncUntrusted}{}
\State $invNum \gets \Call{EnterFunc}{\text{funcID}}$
\State $isRR, isRAW \gets \Call{ExecuteInstrumentedCode}{invNum}$
\State \Call{ExitFunc}{\text{unique funcID}, isRR, isRAW} 
\EndFunction

\Function{FuncTrusted}{invNum: int}
\State $isRR, isRAW \gets \Call{ExecuteInstrumentedCode}{invNum}$ 
\State \textbf{return} $isRR, isRAW$ 
\EndFunction

\end{algorithmic}
\end{algorithm}

\noindent \textbf{Algorithm for Protected Functions.}
Alg.~\ref{alg:instrumented} implements the storage access tracking 
within instrumented protected functions. Table~\ref{tab:instrumentation} 
provides a detailed breakdown of the instrumentation made to the original
functions.
Given an original function implementation, 
\texttt{Func}, it is replicated into two functions: 
\texttt{FuncTrusted} and \texttt{FuncUntrusted}.
\texttt{FuncTrusted} can only be invoked by protected contracts~\footnote{
    \texttt{FuncTrusted} has access control to check whether the \texttt{msg.sender} is 
    in the whitelist of protected contracts, which 
    is not detailed in Alg.~\ref{alg:instrumented} for simplicity.
    }, 
where the \texttt{invNum} is passed by its caller. 
In contrast, \texttt{FuncUntrusted} is designed to
handle invocations from untrusted contracts or external wallets. 
It fetches the \texttt{invNum} from the guard contract 
by calling \texttt{EnterFunc}, tracks the storage accesses, and 
determines whether the function is runtime read-only or involves 
RAW (read-after-write) dependencies before sending the function 
identifier to the guard contract via \texttt{ExitFunc}. Both functions call \\
\texttt{ExecuteInstrumentedCode}, which tracks 
state changes and evaluates whether the invocation is runtime read-only 
and whether any read-after-write dependencies exist. 

Alg.~\ref{alg:instrumented} initializes \texttt{readElements} 
and \texttt{writeElements} 
arrays to store accessed storage slots, alongside two boolean flags, 
\texttt{isRR} (runtime read-only) and \texttt{isRAW} (read-after-write dependencies), 
as described in lines 5-7 of the implementation. 
Then Alg.~\ref{alg:instrumented} proceeds to execute the instrumented code (line 8), 
with specifics provided in Table~\ref{tab:instrumentation}.
After each \texttt{SLOAD} operation, instrumentation appends 
the accessed slot to \texttt{readElements} and records 
it in \texttt{tempReads} if it hasn't previously been written to. 
Correspondingly, each \texttt{SSTORE} operation results in the slot being 
added to \texttt{writeElements} and its value stored in \texttt{tempWrites}. 
If any EVM opcode or function call modifies the blockchain state or 
if a subsequent protected contract call is not runtime read-only, 
\texttt{isRR} is set to \texttt{false}. Additionally, if any subsequent 
call to protected contracts involves a 
read-after-write dependency, \texttt{isRAW} is set to \texttt{true}.

\begin{table}[t]
    \centering
    \caption{Instrumentation for Protected Functions}
    \vspace{-3mm}
    \begin{tabular}{|>{\raggedright\arraybackslash}p{0.4\columnwidth}|>{\raggedright\arraybackslash}p{0.54\columnwidth}|}
    \hline
    \textbf{Original Code} & \textbf{Instrumentation Needed} \\ \hline
    After every SLOAD ($sload(slot) \to value$): & 
    \vspace{-3mm}
    \begin{algorithmic}[1]
        \State $readElements.append(slot)$
        \If{$slot \not\in tempWrites$}
            \State $tempReads[slot] \gets value$
        \EndIf
    \end{algorithmic} 
    \\ \hline
    After every SSTORE ($sstore(slot, value)$): & 
    \vspace{-3mm}
    \begin{algorithmic}[1]
        \State $writeElements.append(slot)$
        \State $tempWrites[slot] \gets value$
    \vspace{-3mm}
    \end{algorithmic} 
    
    \\ \hline
    After other state-changing opcodes or 
    external non-read-only calls: & 
    \vspace{-3mm}
    \begin{algorithmic}[1]
        \State Set $isRR \gets \textbf{false}$
    \end{algorithmic} 
    \vspace{-3mm}
    \\ \hline
    After every call to other protected contracts ($funcCall() \to isSubRR, isSubRAW$): & 
    \vspace{-3mm}
    \begin{algorithmic}[1]
        \State $isRR \gets isRR \land isSubRR$
        \State $isRAW \gets isRAW \lor isSubRAW$
    \end{algorithmic} 
    \\ \hline
    \end{tabular}
    \label{tab:instrumentation}
    \vspace{-5mm}
\end{table}

Finally, Alg.~\ref{alg:instrumented} checks and 
updates \texttt{isRR} and \texttt{isRAW}, 
as well as \texttt{storageWrites} for future 
invocations (lines 9-19). 
For each written storage slot, the slot is recorded in \texttt{storageWrites} 
(line 10). If a slot written is not restored-on-exit,
the function is marked as not runtime read-only(lines 11-12). 
For each slot read, Alg.~\ref{alg:instrumented} checks whether the slot was written 
to in a previous invocation, marking the invocation as RAW-dependent 
if so (lines 13-17).
The temporary mappings are cleared (line 18) before returning 
\texttt{isRR} and \texttt{isRAW} to the guard contract (line 19).

\vspace{-1ex}
\subsection{Gas Optimizations}
\label{subsec:optimization}
\vspace{1sp}

We have implemented \revise{four} optimizations to reduce gas costs.

\noindent \textbf{Optimization 1: Bypassing Validation for Simple Invocations from EOAs. }
When a protected function is not designed 
to invoke arbitrary untrusted contracts given by users(
a prerequisite for re-entrancy attacks),
and 
is directly invoked by an EOA, the transaction will 
consistently follow a single, 
straightforward invocation path, which conforms 
to the criteria set by Policy 1 
(Section~\ref{subsec:policy}). 
We use Slither~\cite{feist2019slither} to 
identify these functions and manually verify to mitigate 
occasional false positives. 
For the verified functions, we insert 
an EOA check at the beginning of execution.
When the caller is an EOA, control-flow validation 
can be safely bypassed, substantially reducing gas overhead.


\noindent
\textbf{Optimization 2: Detecting Restore-on-Exit Storage Slots Statically. }
Another optimization involves statically detecting restore-on-exit storage slots. 
By analyzing a function's control flow graph, we can identify certain storage slots 
that are restored to their original values at the end of every execution branch. 
If such restore-on-exit slots are detected statically, they do not need to be tracked 
at runtime, reducing the overhead of monitoring storage reads and writes. 
We implemented a prototype of this optimization on top of the open-source 
EVM bytecode analysis tool Heimdall~\cite{heimdallrs}. When applied to 
the protected contracts analyzed in Section~\ref{sec:study}, we identified 
$4$ contracts and $50$ functions that utilize re-entrancy guards.

\noindent \textbf{Optimization 3: Merging Guard Contract. }
We merge the guard contract with the most frequently used protected 
contract to convert external \texttt{EnterFunc} and \texttt{ExitFunc} calls 
into internal calls, reducing gas overhead. 
Since it is hard to predict usage patterns before deployment, 
we employ a simple heuristic: after deployment, we merge with the 
contract having the largest bytecode size.
\footnote{For proxy contracts, we use the 
sum of the bytecode sizes of all implementation contracts. 
For contracts implementing the ERC20 token standard~\cite{ERC20}, 
we discount the size contribution of ERC20 functions, 
following Heuristic~1 in Section~\ref{subsec:heuristics}.}
Larger 
contracts typically implement core protocol 
logic and are more likely to be frequently invoked by users.

\noindent \revise{\textbf{Optimization 4: Bypassing Validation for 
Administrator Transactions. }
Transactions originated from administrators are exempted from 
control flow validation under the assumption that protocol 
administrators act benignly. In our evaluation 
in Section~\ref{sec:evaluation}, 
we identify administrators as the deployers of 
protected contracts, allowing their transactions to bypass validation 
entirely. This optimization eliminates gas overhead 
for protocol maintenance operations while 
preserving security against external threats.
}

\section{Evaluation} 
\label{sec:evaluation}

Our evaluation aims to answer the following research questions:

\begin{enumerate}[label=\textbf{RQ\,\arabic*:}, ref={RQ\,\arabic*}]
\setcounter{enumi}{1}
\item How accurately does {\name} stop hack transactions, considering both true positives and false positives?
\item How do various actors, aside from hackers, introduce new control flows, and \revise{what causes false positives in {\name}}?
\item Can informed hackers bypass {\name}?
\item What are the gas overheads of {\name}?
\item How does the performance of {\name} compare to that of the state-of-the-art tool {\tool}?
\end{enumerate}


\begin{table*}[htbp]
    \scriptsize
    \caption{Summary of Benchmarks and Control Flow Analysis Results for Victim DeFi Protocols. }
    \label{tab:studyTable}
    \vspace{-3mm}
\begin{tabular}{|l|l|l|l|l!{\vrule width 1.5pt}c!{\vrule width 1.5pt}l|l|l|l|l|l|l|l|l|l|}
    \hline

\multicolumn{5}{|c!{\vrule width 1.5pt}}{\multirow{2}{*}{Benchmarks}} & \multicolumn{1}{c!{\vrule width 1.5pt}}{RQ1} & \multicolumn{10}{c|}{RQ3} \\   \cline{6-16}
\multicolumn{5}{|c!{\vrule width 1.5pt}}{}                                                 &   \multirow{2}{*}{\makecell{Unique CF \\ in Hack?}}
& \multicolumn{2}{c|}{Total} & \multicolumn{2}{c|}{\#P-Tx} & \multicolumn{2}{c|}{\#S-Tx} & \multicolumn{2}{c|}{\#O-Tx} & \multicolumn{2}{c|}{\#E-Tx} \\  \cline{1-5}  \cline{7-16}

    Victim Protocol                & Protocol Type                        & Hack                               & Hack Type                            & \#C                                  &                        & \#Txs & \#nCF & \#Tx & \#nCF & \#Tx & \#nCF & \#Tx & \#nCF & \#Tx & \#nCF \\  \Xhline{1.5pt}
    bZx                            & Lending                              & \cite{bZx2}                        & oracle manipulation                  & 7                                    & $\halfmark$            & 28712                       & 39                            & 1389                         & 22                           & 16060                         & 18                           & 7729                          & 14                                                    & 3534                          & 14                            \\
\hline
    \end{tabular}
    \vspace{-4mm}
\end{table*}

\vspace{-1ex}
\subsection{Methodology}
\label{subsec:study_methodology}

\noindent \textbf{Hacked Protocol Selection:}
We systematically selected hacked DeFi protocols that experienced 
significant financial losses (exceeding \$300k) on Ethereum. 
Protocols were selected from three complementary sources:
(1) victim protocols identified by~\cite{chen2024demystifying} 
between February 14, 2020 and August 1, 2022; 
\revise{(2) hacked protocols reported by~\cite{zhang2023your} 
within the same period;} and 
(3) protocols compromised between February and July 2024, 
as documented by DeFiHackLabs~\cite{DeFiHackLabs}. 
Cross-chain bridge hacks were excluded.

\noindent \textbf{Target Contract Selection and Protocol Filtering:}
For each selected protocol, we identified all relevant victim protocol contracts 
by analyzing deployer addresses and labels on Etherscan~\cite{Etherscan}. 
We then manually examined the control flow 
of each hack transaction with respect to these contracts. 
\revise{We filtered out hacks involving offchain components and 
hacks with only trivial control flows w.r.t. the protocol contracts, 
in line with our study's scope (see Section~\ref{sec:preli}).
After this filtering process, our final dataset consists of 
$37$ hack incidents: $21$ from source (1), $8$ from source (2), and $8$ from source (3).}
Table~\ref{tab:studyTable} presents our selected benchmarks, 
including their \textit{Protocol Type}, hack transaction links(\textit{Hack}), exploited
 vulnerability category (\textit{Hack Type}), and number of 
 affected contracts (\textit{\#C}). \revise{The vulnerabilities 
 span diverse attack vectors including oracle manipulation,
re-entrancy, DAO governance attack, access control failures,
and input validation errors. Our selected protocols represent a 
comprehensive spectrum of DeFi categories, ensuring broad 
applicability of our findings.}

\noindent \textbf{Transaction History Retrieval:}
We then retrieved the transaction history of these 
identified contracts from their deployment until the hack. 
This complete transaction history, including the hack transaction itself, 
constitutes the dataset of each victim protocol.~\footnote{Although we 
might miss a few affected protocol contracts due to varying deployers or indirect 
involvement, having more contracts to protect would only introduce more complexity 
to the control flows, not reduce it. Thus, our analysis remains valid even with a 
subset of core contracts.} Then we evaluate
{\name} on this dataset.


\noindent \textbf{{\name} Evaluation(RQ2):}
To evaluate the effectiveness of {\name}, we conducted 
experiments under four configurations:
(1) \textbf{Baseline}: A prototype implementing only whitelisting policies 1 and 2 
(see Section~\ref{subsec:heuristics}).
(2) \textbf{Baseline+RR}: Baseline augmented with Policy 3.
(3) \textbf{Baseline+RR+RE}: Baseline augmented with Policy 3 and 4.
(4) \textbf{Baseline+RR+RE+ERC20}: Baseline augmented with Policy 3 and 4, and Heuristic 1.
(5) \textbf{{\name}}: Integrates all 4 policies and 2 heuristics into the Baseline.
Note that these configurations are instrumented pre-deployment and operate autonomously 
post-deployment without manual intervention.
{\name} enforces predefined policies and heuristics without relying on past transaction 
data. However, if an unseen control flow 
is mistakenly blocked, {\name} provides an administrative feedback mechanism that 
allows protocol administrators to manually 
approve and whitelist it. To evaluate this mechanism, we tested {\name} under three {\name}+Feedback settings, 
assuming administrators could approve new control flows within 3 days (19,200 blocks), 
1 day (6,400 blocks), and 1 hour (267 blocks).

We assessed these configurations using historical transactions from \revise{$37$} benchmarks 
collected in Section~\ref{sec:study}.
To further evaluate {\name} under extreme conditions, we applied it to another $3$ widely 
adopted DeFi protocols(AAVE, Lido, and Uniswap) which serve as fundamental 
DeFi building blocks. 
These protocols attract many DeFi developers and feature the most complex and 
continuously evolving control flows due to their high composability and extensive integrations.
To conduct this evaluation, we collected the core smart contracts for these protocols 
from their official websites~\cite{aave2024, lido2024, uniswap2024}. 
Next, we retrieved the most recent 100,000 transactions interacting with these contracts. 
We then applied {\name} to these transactions, measuring its false positive rate (FP\%) 
under real-world extreme conditions. The experimental results are summarized in 
Table~\ref{tab:ablation}.

\noindent \textbf{DeFi Actors Identification(RQ3):}
To deeply understand the results of {\name} and the diversity of control flows in DeFi protocols,
we categorize transactions according to their origins and 
initiators. 
We focus explicitly on transactions with non-trivial control flows (nCFs), as these represent complex and less predictable interactions, offering deeper insights into protocol dynamics. We identify four primary actor groups capable of introducing unique, non-trivial control flows:
\textbf{1. Privileged Transactions (P-Tx)}: Originated by protocol deployers or administrators via privileged functions (e.g., constructors, administrative operations), typically reflecting protocol setup or administrative management activities.
\textbf{2. Same Protocol (S-Tx)}: Transactions initiated by other contracts within the same protocol, developed internally to enhance operational coherence and overall functionality. 
\textbf{3. Other DeFi Protocols (O-Tx)}: Transactions initiated by externally deployed DeFi protocols (commonly labeled on Etherscan), often through open-source collaboration, enriching the broader DeFi ecosystem.
\textbf{4. External Actors (E-Tx)}: Transactions initiated 
by contracts deployed by external actors 
such as arbitrageurs, advanced DeFi users, 
or malicious entities (hackers), generally 
employing closed-source contracts 
to execute complex 
strategies or exploit vulnerabilities.

\vspace{-1ex}
\subsection{RQ2: Effectiveness of {\name}}
\label{subsec:rq2}
\vspace{1sp}



\begin{table*}[htbp]
    \scriptsize
    \caption{Ablation study, Gas Consumption and Bypassability of {\name}}
    \label{tab:ablation}
    \vspace{-3mm}
\begin{tabular}{|l|cc|cc|cc|cc|ccc|cccc|c|c|}   

        \hline
        \rowcolor[HTML]{D9D9D9}
        \multicolumn{11}{|c|}{RQ2} & \multicolumn{1}{c|}{RQ5} & \multicolumn{4}{c|}{RQ2} & \multicolumn{2}{c|}{RQ4} \\ \hline

                                                                        & \multicolumn{2}{c|}{}                                                                                                                                   & \multicolumn{2}{c|}{\multirow{2}{*}{\makecell{Baseline\\+RR}}}                                             & \multicolumn{2}{c|}{\multirow{2}{*}{\makecell{Baseline\\+RR+RE}}}                                        & \multicolumn{2}{c|}{\multirow{2}{*}{\makecell{Baseline\\+RR+RE+ERC20}}}                                   & \multicolumn{3}{c|}{\multirow{2}{*}{\makecell{Baseline+RR+RE+ERC20\\+RAW (a.k.a, CrossGuard)}}}                                                                                                        & \multicolumn{4}{c|}{CrossGuard+Feedback}                                                                                                                                                                                                                                                                                              & \multicolumn{1}{l|}{}                                                 &  \multicolumn{1}{l|}{}                                           \\
\hline

\end{tabular}
\vspace{-4mm}
\end{table*}

The columns ``RQ2'' in Table~\ref{tab:ablation} present the results for RQ2. 
Each configuration is evaluated using two key metrics: ``Block?'' indicates whether the 
hack was successfully blocked; ``FP\%'' represents the 
false positive rate for that configuration. Two sets of benchmarks, $37$ hacked protocols and 
$3$ popular protocols, both include a ``Summary'' row at the bottom, showing the total number of 
blocked hacks and the average false positive rate per protocol. 
When the ERC20 and RAW heuristics are enabled, {\name} blocks $35$ out of $37$ hacks. 
The only exceptions are Bedrock\_DeFi and Auctus. In the case of Bedrock\_DeFi, 
the attacker exploited a missing input validation vulnerability by 
invoking a single function; while two ERC20 functions were also called, 
they were not essential to the core exploit mechanism. 
Similarly, for Auctus, the attacker exploited an access control 
vulnerability by repeatedly calling the same function 
to drain funds. In both instances, these operations could have been 
executed as independent transactions without relying on the 
complex control flows that {\name} is designed to detect. 
Consequently, {\name} did not block these transactions. 
Overall, for the $35$ blocked attacks, the 
exploits involved intricate control flows that were effectively captured by 
{\name}, demonstrating its robustness against sophisticated hacks.



\vspace{-1ex}
\subsection{RQ3: Control Flows Introduced by Different Actors and False Positives}
\label{subsec:rq3}
\vspace{1sp}


The columns labeled \emph{RQ3} in 
Table~\ref{tab:studyTable} summarize our analysis of non-trivial 
control flows introduced by each transaction category. 
The last row provides the average ratios of transactions 
and control flows for each transaction category.

A key insight from this analysis is that a significant number 
of protocols (\revise{$27$}, as highlighted in gray in the \emph{Total} column) exhibit 
a relatively 
low number of non-trivial control flows ($\leq 9$) throughout 
their operational lifetimes prior to being hacked.
Notably, in \revise{$17$} protocols 
(Warp, CheeseBank, XCarnival, Harvest, 
ValueDeFi, VisorFi, Eminence, IndexFi, 
RevestFi, Punk, DoughFina,
BlueberryProtocol, PikeFinance, GFOX,
OmniNFT, Auctus, MonoXFi), 
the hack was the first external non-trivial control 
flow introduced, 
beyond those generated internally (as indicated by gray cells showing 
$0$ nCF from O-Tx but exactly $1$ nCF from E-Tx)~\footnote{ Two 
exceptions are XCarnival and Harvest. They 
were hacked in multiple hack transactions which 
introduced 3 and 2 non-trivial control flows, respectively.}. 
This insight suggests that 
\revise{$17$} hacks could be simply prevented 
with zero false positive rates 
by restricting \textbf{ALL} external 
(``non-trusted'') developers, allowing 
only the protocol deployers to 
create new control flows.
The analysis reveals that E-Tx and P-Tx are significant sources 
of non-trivial control flows. Although external transactions 
account for only $13.17\%$ of the total, they introduce $65.64\%$ of non-trivial 
control flows, often bringing unexpected interactions and 
potential vulnerabilities. 
In contrast, S-Tx, despite comprising $70.14\%$ of the total, 
contribute to only $20.01\%$ of non-trivial flows, underscoring the fact 
that most users interact directly with the protocol rather than through
other intermediate contracts.

\revise{
    We also utilize the above identified transaction categories to conduct 
    a deeper analysis of false positives in RQ2. We examined the three 
    protocols with the highest false positive rates: bZx2 (3.57\%), 
    UwULend (2.38\%), and 
    RariCapital1 (1.22\%)—all of which are lending protocols. 
    Our investigation reveals that false positives predominantly 
    originate from two specific actor categories: Other DeFi Protocols 
    (O-Tx) accounting for 827 out of 1025 false positives in bZx2, 
    3 out of 468 in UwULend, and 74 out of 87 in RariCapital1; and 
    External Actors (E-Tx) responsible for 198 out of 1025 in bZx2, 
    465 out of 468 in UwULend, and 13 out of 87 in RariCapital1. 
    Additionally, we identified that 9 out of 13 false positives in 
    RariCapital1 stem from MEV operations.}

\revise{
    Further investigation into the nature of these false 
    positives reveals two primary categories 
    of legitimate use cases that introduce novel control flows. 
    The first category involves helper contracts designed to 
    streamline user operations, 
    such as a contract that facilitates depositing three different 
    tokens into UwULend by sequentially invoking the deposit 
    function three times within a single transaction. 
    The second category encompasses innovative functionality extensions, 
    exemplified by patterns observed in UwULend where users deploy 
    contracts that invoke deposit and borrow functions multiple times 
    in sophisticated sequences to maximize their borrowing 
    power and optimize capital efficiency. These legitimate but complex 
    interaction patterns highlight opportunities for future 
    refinement of control 
    flow policies to better accommodate common DeFi usage patterns while 
    maintaining security guarantees.
}

\vspace{-1ex}
\subsection{RQ4 \& 5: Bypassability and Gas Overheads}
\label{subsec:rq4-5}
\vspace{1sp}

\noindent \textbf{Case Studies.}
Given that {\name} operates as a 
fully on-chain runtime system, it is transparent, 
allowing attackers to study its implementations and whitelisted 
control flows. A prevalent concern is whether 
informed attackers could bypass {\name} by splitting complex 
hack transactions into simpler ones. To address this, 
we perform in-depth studies of 
the $35$ hacks blocked by {\name}. Our analysis involves 
scrutinizing the control flows, 
underlying vulnerabilities, and the potential outcomes if attackers were 
to split their transactions.

\noindent \textbf{Results for Case Studies.} 
The \emph{RQ4} column in Table~\ref{tab:ablation} summarizes our findings: 
$31$ out of $35$ 
hacks cannot be bypassed by attackers. 
Specifically, $18$ hacks inherently require complex control flows to 
exploit vulnerabilities (marked as \cmark). 

Additionally, $13$ hacks rely on executing multiple capital-intensive 
functions 
to carry out the exploit. Historically, these attacks have used flash loans, which 
require all steps to be completed within a single transaction. Without flash loans, 
the attackers 
have to risk their own capital while competing with arbitrage bots, 
a scenario we deem as non-bypassable due to the high financial risks (marked as \cmark$\ast$). 
Only five hacks Punk, DoughFina, PikeFinance, Audius, and Auctus 
(marked as \xmark) show a bypass chance for attackers. 
The root causes of these exploits are access control, missing input validation, and 
initialization bugs. Each of these vulnerabilities can be exploited by invoking 
a single function without requiring complex control flows. 
These attacks were initially caught by {\name} because the hackers 
included additional preparatory operations in their transactions. 
However, these preparatory steps can indeed be split and executed separately in separate 
transactions, 
allowing attackers to bypass {\name}.


\noindent \textbf{Experiment.}
To measure the gas overhead of {\name}, we instrumented smart contracts in 
a template-based manner. We insert specific code snippets at key points 
such as 
function entry and exit, as well as during storage access operations. These snippets 
execute at runtime to capture the additional gas consumption introduced by {\name}. We used 
Foundry~\cite{Foundry} to compare the gas costs between the original and instrumented contracts, 
recording overhead differences in various execution phases. This process involved detailed 
assessments of each instrumentation type, functions within the guard contract, and EOA
checks. During transaction replays, we logged the additional gas 
costs incurred at these key execution points to compute the overall gas overhead.

\revise{
	To benchmark against industry standards, we evaluated 
	the Hyperithm protocol~\cite{hyperithm, HyperithmSpherex1, 
	HyperithmSpherex2, HyperithmSpherex3} deployed on May 22, 2025, 
	which is protected by SphereX,
	an industry control flow restriction solution discussed in 
	Section~\ref{sec:related}. We measured the 
	additional gas costs during transaction replays to 
	compute overhead for both \name and 
	SphereX implementations on identical transaction history.
}

\noindent \textbf{Results for Gas Overheads.} 
The RQ5 column in Table~\ref{tab:ablation} presents 
the gas overhead introduced 
by {\name} for each benchmark. On average, the overall gas 
overhead is $3.53\%$. 
Notably, $28$ protocols exhibit a gas overhead below $5\%$, primarily due 
to the high proportion of EOA transactions within these benchmarks. Since 
EOA transactions benefit from Optimization 1 (Section~\ref{subsec:optimization}), 
which reduces unnecessary gas consumption, the resulting gas overhead 
remains minimal.
Even for the three widely used DeFi protocols, which feature a significant 
number of contract-initiated transactions, the average gas overhead 
remains $7.52\%$. This is a reasonable tradeoff considering the 
strong security guarantees provided by {\name}. The higher 
overhead in these cases stems from the need to track function 
calls and storage accesses, which are essential for securing 
protocols with complex execution flows. 

\revise{
Our industry comparison reveals that SphereX introduces 6.09\% 
overhead for the Hyperithm protocol, while {\name} achieves 
only 0.22\% overhead on the same contracts. 
Analysis of 166 transactions shows that 127 are simple 
invocations optimized by Optimization 1, and 38 are 
deployer-originated transactions optimized by Optimization 4. 
In contrast, SphereX lacks these optimizations and triggers 
external calls for every function invocation, 
resulting in substantially higher gas overhead. 
This demonstrates {\name}'s superior efficiency 
through targeted optimization strategies.
}





\vspace{-1ex}
\subsection{RQ6: Comparative Analysis with {\tool}}
\label{subsec:rq6}
\vspace{1sp}

\noindent \textbf{Experiment.}
We compare {\name} against {\tool}~\cite{chen2024demystifying}. {\tool} 
relies on historical transaction data to generate invariants, which are then 
instrumented into smart contracts to prevent hacks. This approach requires 
a training set (TS) of past transactions to learn security rules. In contrast, 
{\name} operates without historical data, making it applicable to new contracts 
before deployment. Additionally, {\name} allows protocol administrators 
to explicitly whitelist control flows they deem safe.
To evaluate their effectiveness, we apply both tools to our $37$ benchmark 
protocols. As required by {\tool}, we use 70\% of transaction history 
as the training set and evaluate both tools on the remaining 30\% of transactions 
as the testing set. We assess two versions of {\name}: 
one operating without training data and another that assumes 
all control flows from the training set are whitelisted. 
We compare {\name} against {\tool} using its two most effective 
security invariants: EOA$\land$GC$\land$DFU and 
EOA$\land$(OB$\lor$DFU)~\cite{chen2024demystifying}.

\vspace{-2mm}
\begin{table}[htbp]
\scriptsize
\caption{Comparison of {\name} and {\tool}}
\label{tab:comparison}
\vspace{-3mm}
\begin{tabular}{|l|c|c|cc|}
\hline
                                            & \multirow{2}{*}{\begin{tabular}[c]{@{}c@{}}CrossGuard \\ (w/o TS)\end{tabular}} & \multirow{2}{*}{\begin{tabular}[c]{@{}c@{}}CrossGuard \\ (w TS)\end{tabular}} & \multicolumn{2}{c|}{Trace2Inv (w TS)}                  \\ \cline{4-5}
                                            &                                                                                      &                                                                                               & \multicolumn{1}{c|}{EOA$\land$GC$\land$DFU} & EOA$\land$(OB$\lor$DFU)                                                                                             \\ \hline
\multicolumn{1}{|c|}{\# Hacks Blocked}      & \revisionhl{35}                                                                                   & \revisionhl{35}                                                                                            & \multicolumn{1}{c|}{\revisionhl{34}}                     & \revisionhl{29}                                                                                                             \\ \hline
\multicolumn{1}{|c|}{Avg. FP\%}             & \revisionhl{1.19}                                                                                 & \revisionhl{0.15}                                                                                          & \multicolumn{1}{c|}{\revisionhl{3.14}}                   & \revisionhl{0.23}                                                                                                           \\ \hline
\end{tabular}
\vspace{-3mm}
\end{table}

\noindent \textbf{Results.} Table~\ref{tab:comparison} presents the analysis results 
that demonstrate that {\name}, even without training data, effectively blocks 
$35$ out of $37$ 
while maintaining an average FP\% rate of $1.19$\%. This is a significant achievement 
compared to {\tool}, which no only requires training data to function but only blocks at most 
$34$ hacks. When trained, {\name} maintains 
its effectiveness in blocking hacks, reducing its FP\% rate dramatically 
to $0.15$\%, which is superior to the FP\% rates achieved by {\tool}'s invariants 
($3.14$\% and $0.23$\%). 
These results underline {\name}'s potential as a state-of-the-art solution providing 
robust security for DeFi applications. Moreover, {\name} and {\tool} can be used in 
conjunction to provide a more comprehensive security solution.


\vspace{-1mm}
\section{Discussion and Threats to Validity}
\label{sec:discussion}

\noindent \textbf{Generalization.} 
{\name} can generalize through two mechanisms. 
First, it can evolve and reduce false positives 
by learning 
from blocked benign transactions, 
enabling administrators to whitelist new control 
flow patterns. 
Second, it records all non-trivial control flows 
in the guard contract. 
If novel attacks bypass current conditions 
(Alg.~\ref{alg:enterexit}, Line 27), 
developers can update the 
guard contract to add new detection logic and automatically revert 
such transactions.
This framework is extensible to track additional runtime 
properties for emerging threats.

\noindent \textbf{Integrating {\name} Pre- and Post-deployment.}
For new protocols, {\name}'s integration involves three steps: 
(1) instrument contracts (Section~\ref{sec:approach}), 
using Slither and Heimdall  
to omit unnecessary instrumentation (Section~\ref{subsec:optimization}); 
(2) deploy a guard contract to enforce policies and heuristics; 
and (3) configure access controls for all components. 
For existing deployed upgradable protocols~\cite{bodell2023proxy}, 
{\name} can be retrofitted 
by upgrading implementation contracts to new 
instrumented versions, 
thereby enabling {\name} without disrupting existing 
protocol logic.
Following deployment, {\name} operates autonomously. 
Administrators can further improve {\name} by 
whitelisting blocked legitimate control flows (false positives).
This adaptive process ensures that {\name} evolves alongside protocol growth,
maintaining strong security guarantees over time.






\vspace{3pt}
\noindent \textbf{Threats to Validity.}
The internal threat to validity concerns potential human errors 
in identifying protected contracts. As discussed in 
Section~\ref{subsec:study_methodology}, 
we rely on Etherscan labels to 
identify these core contracts to protect. The labels 
might be incomplete, leading to missing protected contracts. 
However, with more protected contracts identified, our approach will only
become more effective 
in blocking hacks, as the control flows of hack transactions recorded by 
{\name}
will be more complex but still unique.
Our results may also face external threats due to the reliance on 
Trace2Inv~\cite{chen2024demystifying} and Sting~\cite{zhang2023your} benchmarks, 
which focus on hacks up to 2022. We mitigate this threat by 
including additional 8 hacks 
from February to July 2024. Additionally, we also 
include three major DeFi protocols representing
the most current protocols and user transactions.

\vspace{-1mm}
\section{Related Works} \label{sec:related}

\noindent
\textbf{Invariant Generation and Enforcement.}
Prior research generates and enforces invariants for contract security.
Cider~\cite{liu2022learning} derives arithmetic overflow invariants 
via deep reinforcement learning, 
while InvCon~\cite{liu2022invcon} and InvCon+~\cite{liu2024automated} 
combine dynamic inference with static verification for function-level invariants. 
Trace2Inv~\cite{chen2024demystifying} learns invariants from transaction 
history to prevent attacks. 
Unlike these approaches, which focus on individual 
contracts or specific bugs, {\name} is a general-purpose 
framework that
targets control 
flows across multiple protocol contracts with a one-time 
pre-deployment configuration.

\noindent
\textbf{Control Flow Restriction.}
SphereX~\cite{spherex} in industry offers services to manually restrict control flows, 
requiring developers to explicitly whitelist or 
blacklist paths. 
In contrast, {\name} automates the whitelisting of unseen control 
flows and simplifies the overall structure, 
significantly reducing developer burden and increasing 
system adaptability compared to manual intervention.

\noindent
\textbf{Re-entrancy Attack Defense and Secure Type System.}
Numerous tools restrict control flows to combat re-entrancy attacks. 
Static analysis tools~\cite{luu2016making,torres2018osiris,liu2018reguard,feist2019slither,
zhang2019mpro,bose2022sailfish,ma2021pluto,zheng2022park,wang2024efficiently,
albert2020taming} identify re-entrancy vulnerabilities and apply guards. 
Runtime frameworks like Sereum~\cite{rodler2018sereum} and 
\citeauthor{grossman2017online} protect deployed contracts, 
while \citeauthor{callens2024temporarily} prevent duplicate 
function calls within transactions. Unlike these 
re-entrancy-focused approaches, {\name} adopts broader 
control flow restriction targeting comprehensive vulnerability classes.
\citeauthor{cecchetti2020securing} propose a security type system to 
enforce information flow and re-entrancy controls~\cite{cecchetti2020securing,
cecchetti2021compositional}. Conversely, 
{\name} is a purely runtime system built entirely on the EVM, 
operating without supplementary type systems or language modifications.



\vspace{-1mm}
\section{Conclusion} 
\label{sec:conclusion}

In this paper, we presented {\name}, a novel control 
flow integrity framework
specifically designed to secure DeFi smart contracts at 
runtime. Configured only once at deployment, 
{\name} prevents
malicious transactions from executing risky control flows on the fly,
effectively mitigating a wide range of attacks. 
Our comprehensive evaluation demonstrates that 
{\name} blocks the vast majority of
benchmark attacks, significantly 
reduces false positives without relying on a
pre-collected training set of benign transactions, 
and maintains a manageable
gas overhead. Furthermore, integrating manual 
feedback enhances its accuracy. Together, 
these results establish
{\name} as a practical solution for securing smart contracts.


\begin{acks}
    \noindent 
    We thank anonymous reviewers for their insightful comments on
the early version of the paper.
This work was supported by Mitacs
through the Mitacs Accelerate program.
\end{acks}


\newpage
\bibliographystyle{ACM-Reference-Format}
\bibliography{main}



\end{document}